# Multiple Access Demodulation in the Lifted Signal Graph with Spatial Coupling

Christian Schlegel and Dmitri Truhachev

*Abstract*—Demodulation in a random multiple access channel is considered where the signals are chosen uniformly randomly with unit energy, a model applicable to several modern transmission systems. It is shown that by lifting (replicating) the graph of this system and randomizing the graph connections, a simple iterative cancellation demodulator can be constructed which achieves the same performance as an optimal symbol-by-symbol detector of the original system. The iterative detector has a complexity that is linear in the number of users, while the direct optimal approach is known to be NP-hard. However, the maximal system load of this lifted graph is limited to $\alpha < 2.07$, even for signal-to-noise ratios going to infinity – the system is interference limited. We then show that by introducing spatial coupling between subsequent lifted graphs, and anchoring the initial graphs, this limitation can be avoided and arbitrary system loads are achievable. Our results apply to several well-documented system proposals, such as IDMA, partitioned spreading, and certain forms of MIMO communications.

*Index Terms*—random signaling, iterative decoding, optimal joint detection

## I. INTRODUCTION

ITERATIVE graph-based signal processing has enjoyed a tremendous rise in popularity with the introduction of turbo coding [2]. By breaking complex algorithms into local processes which communicate in iterative cycles, many highly complex problems can now be addressed with relative ease, and the iterative process reduces complexity in a way similar to how iterative solution methods break down the problem of solving large systems of equations into a repetition of operations with much lower complexity [21].

The analysis of the performance of iterative processors, on the other hand, has offered more resistance, and relatively little is known about the exact performance of these algorithms. The methods introduced by ten Brink [30], Divsalar [6], and Richardson et. al. [19] are statistical methods to predict the average dynamical behavior of large iterative error control decoders. These analysis methods have much in common with Gallager's original probability propagation approach used to study low-density parity-check codes [9]. The same type of analysis has been utilized by others [1], [25], [3] to study the behavior of iterative decoders for multiple access systems, which were introduced shortly after the invention of the turbo decoding algorithm [15], [17], [1]. All of these methods can predict the dynamical behavior of iterative processors for large-scale systems with remarkable accuracy, however, they cannot predict exact performance in terms of error rates, etc.

The question how close iterative processors, in particular decoders and demodulators, can approach the performance of an optimal decoder is also not well understood. While it appears that the turbo decoding algorithm has a performance very close to that of a maximum-likelihood processor, it is also know that it is not equivalent to maximum likelihood [14].

In this paper we present a case where an augmented graph-based iterative processor achieves the optimal maximum-likelihood performance of the underlying original system in terms of error rates (and LLR statistics). The system in question is a multiple access system where the waveforms used for modulation are uniformly randomly selected from the $N$–1-dimensional unit(-energy) sphere in $N$-dimension. This model has practical counterparts in random CDMA (code-division multiple-access) as explored in partitioned CDMA [22], [23], interleave-division multiple access (IDMA) [16], or in isotropic multiple antenna communications channel models.

The performance of a maximum-likelihood detector for random CDMA was calculated in [28] where a statistical mechanics approach was used. In this paper we show that a proper augmentation of the factor graph of the random-code model leads to a system in which a basic message passing algorithm can be applied which achieves the same performance as the optimal detector in the original system as computed in [28].

However, from a performance standpoint, the iterative system is ultimately interference limited in the sense that the maximal ratio of signals per dimensions $\alpha = K/N < 2.07$ in an equal received power situation. We then show that this interference limitation can be overcome by a process called *spatial coupling*, whereby successive frames of the original system are interconnected, and the first few frames are initiated with known symbols which act as anchor or pilot symbols. We show that with appropriate such anchoring the interference limitation disappears and the supportable loads $\alpha \to \infty$, given sufficient signal-to-noise ratio.

## II. SYSTEM MODEL

We consider a general communications system with $K$ data streams. Without essential loss of generality, the data consists of independent binary symbols $d_k \in \{-1, 1\}$, $k = 1, \ldots, K$ which may be the outputs of $K$ parallel forward error (FEC) control encoders. The binary symbols are then modulated onto random unit-energy waveforms $\mathbf{a}_k$ chosen uniformly from the $N$–1 dimensional unit sphere in $N$ dimensions – see Appendix II for the signal description.

Christian Schlegel and Dmitri Truhachev are with the Department of Computing Science, University of Alberta, Edmonton, Canada.
Supported in part by *iCORE* Alberta.

The composite signal after transmission at the receiver is given by

$$\mathbf{y} = \sum_{k=1}^{K} d_k \mathbf{a}_k + \sigma \mathbf{n} \quad (1)$$

where $\mathbf{n}$ is a vector of unit energy noise samples, and $\sigma$ is the noise standard deviation. As shown in Appendix II, the average cross-correlation of these waveforms is $E[\mathbf{a}_k^T \mathbf{a}_j] = 1/N, k \neq j$, which is all we need[1].

In the sequel we will use graph-based arguments and we therefore introduce the "factor graph" [11], [10] of (1), shown in Figure 1, where the center node symbolizes the addition of the $K$ signals.

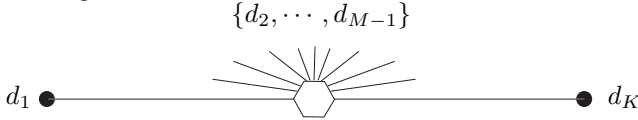

Fig. 1. Factor graph of the basic multiple-access system of Equation 1.

A decoder, or demodulator, can utilize this graph as an algorithmic blueprint. In this case the graph is not very interesting, and consists of a single, cycle-free star graph. An (optimal) demodulator will have to compute

$$\hat{\mathbf{d}} = \arg\min_{\mathbf{d}} \|\mathbf{y} - \sum_{k=1}^{K} d_k \mathbf{a}_k\|^2 \quad (2)$$

which involves the computation of $2^K$ terms. In any case, it is known that (2) is NP-hard [34], and there is little hope that efficient algorithms for its exact solution exist unless P=NP, which most experts agree is not likely the case [8].

Even though the general problem is NP-complete, special instances of the multiple access problem (2) have efficient, low-complexity solution algorithms. For example, if $\mathbf{a}_k^T \mathbf{a}_l = 0, l \neq k$, which is usually done by design, see e.g. [33], demodulation complexity is of order $O(NK)$ only. If the cross-correlations between the different signals take on only nonpositive values, optimal detection corresponds to a solution of a min-cut problem of the associated graph, whose complexity is $O(NK^3)$ [20], [32]. Sequences $\mathbf{a}_k$ can also be constructed specifically such that lower-complexity optimal detection is possible, even when the system is over-saturated, i.e., $\alpha = K/N > 1$ [13], where $\alpha$ is called the *system load*. In [24] the authors introduce controlled additional sequences such that the corresponding over-saturated system can be demodulated with a trellis decoder, whose complexity depends on $\alpha$. All these approaches, of course, are only feasible if timing of the different signals can be tightly controlled and no transmission distortions occur.

Like the optimal sequence detector, the symbol-wise marginal-posterior-mode (MPM) detector, which maximizes the symbol-wise posteriori probability, is NP-hard in general. The probability of error for the MPM detector was computed in [28] using statistical mechanics tools. The signal-to-interference ratio $\gamma^2$ of the posterior probabilities of $d_k$ prior to thresholding is computed as the solution to the implicit equation ([28, Eqns. 45 and 43])

$$\gamma^2 = \left[\sigma^2 + \alpha E\left[1 - \tanh\left(\frac{F}{E}\left(\gamma^2 + \gamma\xi\right)\right)\right]^2\right]^{-1} \quad (3)$$

where $\xi \sim \mathcal{N}(0,1)$, and $E$ and $F$ are given in [28, Eqn. 43], and are the mean and variance of the (individual) APP output signal. In general $E > F$, however in the case of binary random signaling as used in this paper $E = F$, which follows from the definitions of $m, q, E, F$ in [28, Eqn. 27 ff.]. The result (3) holds for loads $\alpha < \alpha_s \approx 1.49$, where $\alpha_s$ is the "spinodal value", where the statistical mechanics approach undergoes a phase transition and multiple solutions appear.

Mathematically, the spinodal value separates the region where a single solution exists to (3), from that where multiple solutions exist. More precisely,

$$x - \sigma^2 - \alpha E\left[1 - \tanh\left(\left(\frac{1}{x} + \sqrt{\frac{1}{x}}\xi\right)\right)\right]^2 \quad (4)$$

is monotonically non-decreasing for in $x$ $\alpha < \alpha_s$, and therefore only a single solution to (3) exists. For $\alpha > \alpha_s$ there are multiple solutions, and the iterative detector presented here will converge to the solution with minimum $\gamma$.

In this paper we show that by process called graph lifting and randomization of connection, we can obtain an instance of the decoding problem where an iterative message passing algorithm can achieve a posterior LLR signal-to-interference ratio that is identical to (3), and therefore achieves the same error performance as the MPM detector, but with a decoding complexity of $O(IN)$ per bit $d_k$, where $I$ is the number of iterations. The number of iterations $I$ does not depend on $K$, and in most cases, especially those of practical interest, $O(IN) \leq O(N2^K/K)$.

### III. GRAPH LIFTING

We now consider an equivalent model for the system (1). Imagine that each bit is replilcated $M$ times as $d_{k,m} = d_k, \forall m = 1, \cdots, M$, i.e., the transmission model is

$$\mathbf{y} = \sum_{k=1}^{K} \sum_{m=1}^{M} \frac{d_{k,m}}{M} \mathbf{a}_k + \sigma \mathbf{n} \quad (5)$$

The factor graph of (5) is shown in Figure 2.

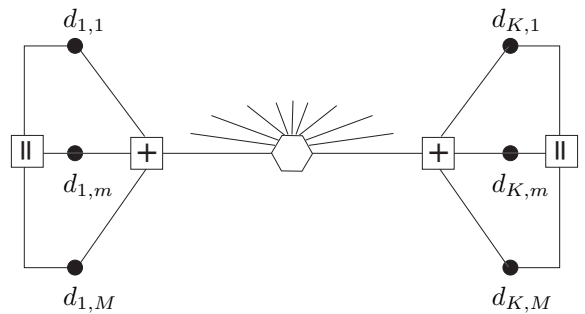

Fig. 2. Factor graph of the basic multiple-access system of Equation 1.

---

[1]Note that the waveforms can, of course, be complex as is the case in typical wireless communications systems.

However, since (5) is linear, we find the representation in Figure 3 more useful. Note that the central multiple access node now has an $M$-fold larger incidence degree.

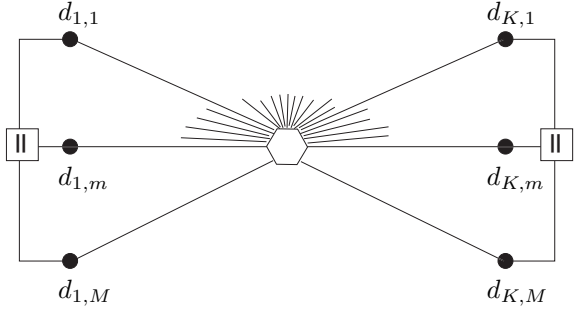

Fig. 3. Factor graph of the basic multiple-access system of Equation 1.

Figure 4 shows an $L$-fold lifting of the graph of Figure 3, that is, basically an $L$-fold repetition. This can be thought of in practice as the transmission of a block of $LM$ signals instead of the original $M$ signals for each user.

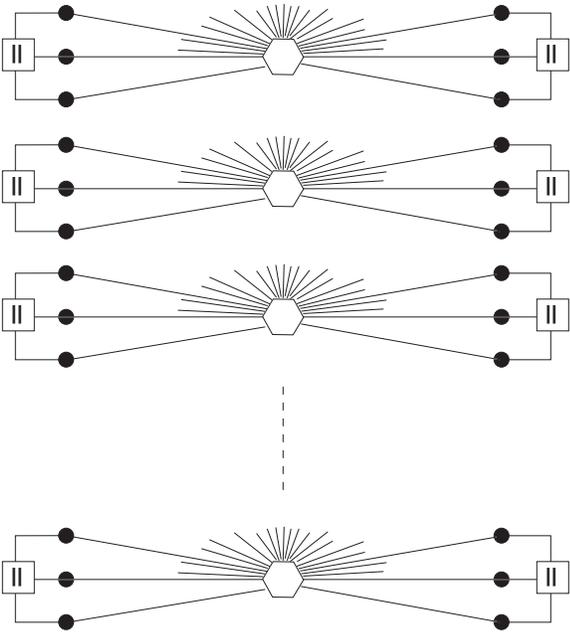

Fig. 4. $L$-fold lifting of the factor graph of the basic multiple-access system of Equation 1.

Of course, up to now nothing has changed in the system or the processing of the data. However, analogously to the construction of LDPC codes from protograph structures [31], [5], a concept initiated by Tanner [29], we now permute the edges among the different copies of the graph lifting. This can be done in a number of convenient ways, such as to avoid contention by a memory-based receiver processor [27]. However, here we simply assume that the permutations are done randomly, and independently from each other. The resulting multiple access equation is given as

$$\mathbf{y}_l = \sum_{k=1}^{K} \sum_{m=1}^{M} \frac{d_{k,m',l'}}{\sqrt{M}} \mathbf{a}_{k,l'} + \sigma \mathbf{n}_l \quad (6)$$

where we have introduced the graph copy indices $l$ and $l'$, and $(m', l')$ are simply the indices of the original location $Ml'+m'$ of an edge that is permuted to location $Ml + m$ in (6). As a small detail we note that in order to keep the total energy of the system constant, we had to change the amplitudes of the waveforms to $1/\sqrt{M}$. This is because the individual portions of a given waveform do no longer add coherently as in (5). Figure 5 shows the resulting graphical model.

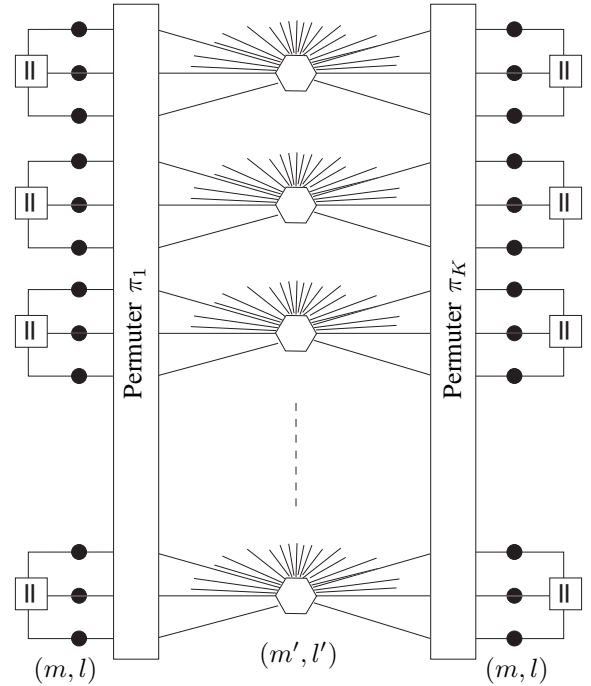

Fig. 5. Graphical structure created from the $L$-fold lifting of the original factor graph by adding individual permuters.

The key observation is, that instead of solving the NP-hard decoding problem for (1), we apply an iterative message passing algorithm to the lifted graph in Figure 5, and we show that this iterative algorithm achieves the same performance as the maximum-likelihood algorithm for the original problem. This is significant in that the exponential complexity per bit has been transformed into the complexity of the message passing algorithm per graph section. The latter, as we will show, only grows linearly in $K$.

## IV. BELIEF PROPAGATION MESSAGE PASSING

Message passing in the lifted graph is carried out in the usual way. The equality nodes compute an extrinsic log-likelihood ratio of the bits $d_{k,m,l}$ as the sum of the incoming a priori log-likelihood ratios, i.e.,

$$\lambda(d_{k,m,l}) = \sum_{n:n\neq m} \lambda(d_{k,n,l}) \quad (7)$$

where $\lambda(d) = \log\left(\Pr(d=1)/\Pr(d=-1)\right)$ formally plays the role of a log-likelihood ratio. The details of the message passing operation are illustrated in Figure 6.

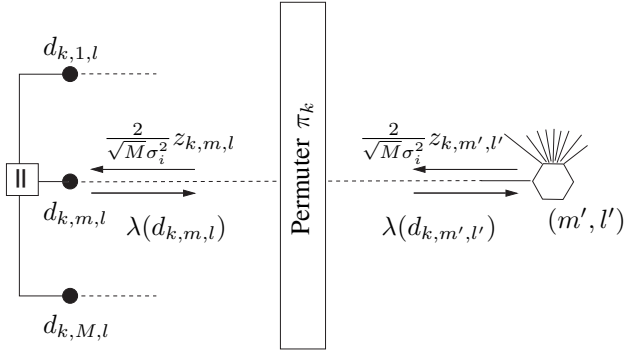

Fig. 6. Illustration of the message exchange mechanism and node functionalities in the message passing algorithm.

Message passing through the multiple access node is a little more involved, and an exact calculation of the extrinsic log-likelihood ratio would lead us back to the NP-hard problem we wanted to avoid in the first place. We therefore utilize a much simpler cancelation and subsequent unconstrained LLR computation. For each outgoing edge from the $l'$-th multiple access node, the following cancelled and filtered signal is computed and sent to the symbol nodes:

$$z_{k,m',l'} = \mathbf{a}_{k,l'}^T \left( \mathbf{y}_{l'} - \sum_{\substack{\kappa,\mu \\ (\kappa,\mu)\neq(k,m')}} \tanh\left(\frac{\lambda(d_{\kappa,\mu,l'})}{2}\right) \frac{\mathbf{a}_{\kappa,l'}}{\sqrt{M}} \right)$$

$$\lambda(d_{k,m',l'}) = \frac{2}{\sqrt{M}\sigma_i^2} z_{k,m',l'} \quad (8)$$

where $\sigma_i^2$ is the joint multiple access and noise interference of the (partially) canceled signal $z_{k,(m',l')}$. In the limit of $K \to \infty$, by virtue of being a sum of i.i.d. bounded random variables $z_{k,(m',l')}$ becomes Gaussian with variance $\sigma_i^2$ and mean $d_{k,(m',l')}/\sqrt{M}$ by the central limit theorem. Evidently, its variance is independent of the lifting index $l$, the user index $k$, and the partition index $m$. The index $i$ can be understood as an iteration index, since this variance changes with repeated iterations of the processing operations (7) and (8) – see also Figure 6.

For the signals used $\mathrm{E}\left[\left(\mathbf{a}_{k,l}^T \mathbf{a}_{m,j}\right)^2\right] = 1/N$, (see Appendix II), and noting that the LLRs in (8) are distributed with mean $2/(M\sigma_i^2)$ and variance $4/(M\sigma_i^2)$, and accounting for (7), it follows that

$$\sigma_i^2 = \frac{1}{NM} \sum_{\substack{\kappa,\mu \\ (\kappa,\mu)\neq(k,m'))}} \mathrm{E}\left[d_{\kappa,(\mu,l')} - \tanh\left(\frac{\lambda(d_{\kappa,(\mu,l')})}{2}\right)\right]^2 + \sigma^2$$

$$= \frac{1}{NM} \sum_{\substack{\kappa,\mu \\ (\kappa,\mu)\neq(k,m'))}} \mathrm{E}\left[1-\tanh\left(\frac{M-1}{M\sigma_{i-1}^2} + \sqrt{\frac{M-1}{M\sigma_{i-1}^2}}\xi\right)\right]^2 + \sigma^2 \quad (9)$$

where $\xi \sim \mathcal{N}(0,1)$. That is, assuming that (8) describes a Gaussian random variable – which is true for $K$ large – (9) is an exact expression for all $i < I_{\text{cycle}}/2$, where $I_{\text{cycle}}$ is the length of the shortest cycle in the graph of Figure 5. Equation (9) is therefore exact for as long as the system size $L$ is large enough to ensure large minimum cycles. Typically this means that $L$ grows as $\propto e^{I_{\text{cycle}}}$. Nonetheless, the analysis doesn't collapse for shorter $L$, but becomes increasingly approximate. We shall assume in what follows that $L \to \infty$, without deriving any practical suggestions from this statement.

Equation (9) is monotonically decreasing due to the monotonicity of the expectation function in (9). It is a contraction map with a unique fixed point according to the Banach fixed point theorem. This fixed point is given implicitly by

$$\sigma_\infty^2 = \frac{KM-1}{NM}\mathrm{E}\left[1-\tanh\left(\frac{M-1}{M\sigma_\infty^2}+\sqrt{\frac{M-1}{M\sigma_\infty^2}}\xi\right)\right]^2 + \sigma^2$$

$$\to \alpha\,\mathrm{E}\left[1-\tanh\left(\frac{M-1}{M\sigma_\infty^2}+\sqrt{\frac{M-1}{M\sigma_\infty^2}}\xi\right)\right]^2 + \sigma^2 \quad (10)$$

where in the second row we have used $KM \gg 1$, which is the case for most situations of practical interest.

Note that the analysis up to this point does not depend on the specific binary sequence chosen in the $K$ data streams, and one can assume the all-zero sequence for all of them, also that is not required.

At each iteration, the equality node will collect the LLR messages reported from the network and form an $i$-th soft symbol estimate (or a posteriori LLR value) to be reported to the external system. This LLR is given by

$$\lambda(d_{k,l}) = \sum_{n=1}^{M} \lambda(d_{k,n,l}) \quad (11)$$

for each binary symbol $d_{k,l}$. Following (6), (10), and (7), the signal-to-noise ratio of this estimate is given by $\gamma_i^2 = 1/\sigma_i^2$, and follows the iteration equation

$$\gamma_i^2 = \left(\alpha\,\mathrm{E}\left[1-\tanh\left(\frac{M-1}{M}\gamma_{i-1}^2+\sqrt{\frac{M-1}{M}}\gamma_{i-1}\xi\right)\right]^2+\sigma^2\right)^{-1} \quad (12)$$

We note that for large partition numbers $M \to \infty$, the fixed point SNR of (12) and the implicit SNR equation for the optimal symbol detector (3) are identical. That is, the SNR $\gamma_i^2$ approaches the optimal symbol SNR as iterations approach the fixed point. This shows that simple message passing in the (large) lifted and randomized graph of the random signal multiple access system achieves optimal performance, i.e., its error performance is identical to that of operating an NP-hard maximum marginal probability detector on the original system (also the detected binary sequences would not necessarily be identical with probability one for any finite signal-to-noise ratio).

## V. DISCUSSION

For $\sigma^2 > \sigma^2_{\text{crit}} = 0.148$, equation (12) has only a single solution. However, for $\sigma^2 \leq \sigma^2_{\text{crit}}$, one, two or three solutions exist, depending both on $\sigma^2$ as well as $\alpha$. The relevant solution, i.e., the fixed point approached by the iterative detector, is the one with the largest $\sigma^2_\infty$. Only the case with a single solution is practically relevant, since the other solutions are high-noise fix points where the decoder essentially fails, i.e., produces high bit error rates (see Figure 7 below). For each $\sigma^2 \leq \sigma^2_{\text{crit}}$, we have an associated maximum $\alpha$ such that a single solution exists. For $\sigma^2 \to 0$, this maximum $\alpha$ is the total maximum system load that is supportable at any SNR, given by $\alpha_{\max} = 2.07425$.

Figure 7 shows how the appearance of additional solutions translates into a "turbo-cliff" behavior of the bit error rate. This abrupt phase change is therefore characteristic of the optimal detector as well. Our results also suggest that the phase transitions observed in the statistical mechanics approach [28, Figure 1] are artifacts of the analysis, and are not a phenomenon of the detector itself, whose error curves transition sharply as the fixed point jumps at the critical SNR.

The bit error rate was computed from the signal-to-noise ratio $\gamma^2_\infty$ as $P_b = Q(\gamma_\infty)$. The dotted line is the bit error rate of antipodal signaling on an AWGN channel. We can deduce, for example, that at $\alpha = 1$ the cost of using random signaling instead of orthogonal signals is about 2dB at $P_b = 10^{-2}$, and shrinks to a small fraction of a dB for lower error rates.

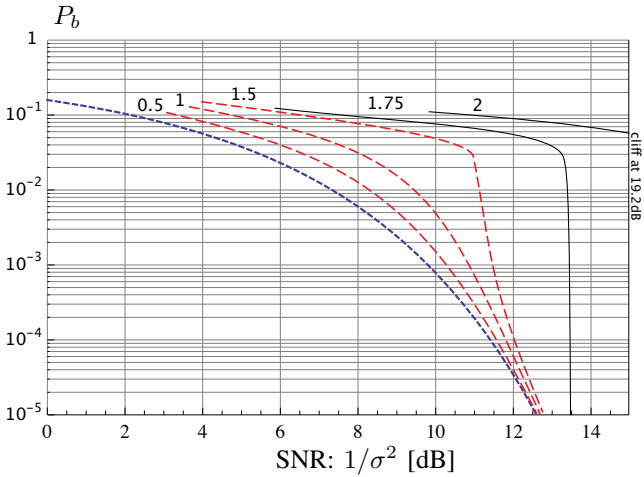

Fig. 7. Bit error rate of the MPM detector for various $\alpha = K/N$.

Just as with irregular low-density parity-check codes, where an optimization of the node degrees led to significant improvements in performance [18], the partition degree of our lifted graph does not have to be equal for all users, or all symbols for that matter. We are free to chose a partition degree profile $M_{1,1}, \cdots, M_{1,K}, M_{2,1}, \cdots, M_{L,K}$. The fact that gains in performance can be achieved can be seen through the following argument. Letting $M_{l,k} = M_k$, it is not hard to see that, as long as each symbol is transmitted with constant power $1/M$, this amounts to assigning different powers to the different users. However, it is already known that with exponentially distributed power assignments, or equivalent rate assignments, the multiple access channel capacity can be achieved with a cancelation system of this type [23].

Lastly, we wish to comment on a subtle difference in the use of the signal space of the lifted system, compared to the original system. As mentioned in Section II (and Appendix II), the signals $\mathbf{a}$ are drawn uniformly and randomly from the $N$-dimensional signal space. With the lifting, each signal is split into $M$ signals, also drawn uniformly randomly. Note that the total power and the total bandwidth utilization are equal in both systems. However, the lifted signal has a higher signal density, as $M$ times as many signal crowd the original signal space. This is illustrated in Figure 8.

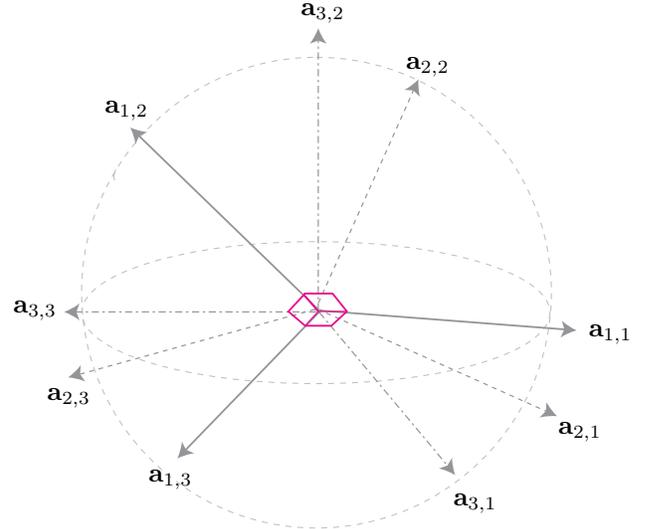

Fig. 8. Illustration of the redundant signal population using the lifted system.

Note that without loss generality to the analysis we may wish to enforce that the signals belonging to the same user $k$, and the same time interval $l$, are chosen to be mutually orthogonal, i.e., $\mathbf{a}_{k,l',1} \perp \mathbf{a}_{k,l',2} \perp \cdots \perp \mathbf{a}_{k,l',M}$. This causes no significant design overhead, since instead of using $M$ random signals, each user now uses a random $M$-dimensional basis to transmit its symbol fragments in time interval $l$. The analysis presented in this paper carries through without change. However, from an operational point of view, signal processing at the receiver is facilitated since no intra-user inter-signal interference needs to be cancelled.

## VI. SPATIAL COUPLING

In this section we are addressing the issue of the maximum system load of $\alpha_{\max}$ introduced in Section V. The question is in what way this is a limit to the iterative detector – in short, it is not. This will be demonstrated in this section.

In the development of LDPC convolutional coding [7], a phenomenon called "spatial coupling" has allowed these codes to be designed with decoding thresholds that are very close to the channel capacity. The effect of spatial coupling derives from anchoring initial symbols to known values, which causes a locally smaller rate. This in turn allows the code to converge in signal-to-noise ratios where uniform convergence is not possible. Recently, it has been shown that spatial coupling can decrease the convergence threshold in low-density parity-check codes on binary-erasure channels all the way to the maximum-likelihood threshold [12].

In our system we introduce spatial coupling by duplicating the graphical structures of Figure 5, i.e., lifting the graph again. One can think of this as repeating signaling in time. We then anchor the signals of the left-most graph to known values, these can be thought of as pilot signals, for example. The spatial coupling in its simplest form is achieved by connection of a fraction $a$ of the nodes of the graph at lifting level (time) $t$, to the multiple access nodes at time $t-1$. This is illustrated in Figure 9, where the solid black nodes denote the anchored known symbols.

By this process, the first set of multiple access nodes effectively experiences a lower load, and will therefore converge faster. Astoundingly, this effect propagates through the semi-infinite lifted graph indefinitely.

In order to simplify the notation and following discussion we define

$$g\left(\frac{M-1}{Mx}\right) = \mathrm{E}\left[1 - \tanh\left(\frac{M-1}{Mx} + \sqrt{\frac{M-1}{Mx}}\xi\right)\right]^2 \quad (13)$$

and we note that $g\left(\frac{M-1}{Mx}\right)$ is a monotonically increasing function in $x$ [4]. We also assume without essential loss of generality that $M$ is large, so we can approximate $M-1 \approx M$.

For the further discussion we substitute $x_i^t = \sigma_{t,i}^2$. A density evolution analysis for large graphs then links the variances at the different times via the following equation.

$$x_i^1 = \alpha a g\left(\frac{1-a}{x_{i-1}^2} + \frac{a}{x_{i-1}^1}\right) + \sigma^2 \quad (14)$$

$$x_i^t = \alpha a g\left(\frac{1-a}{x_{i-1}^{t+1}} + \frac{a}{x_{i-1}^t}\right) + \alpha(1-a) g\left(\frac{1-a}{x_{i-1}^t} + \frac{a}{x_{i-1}^{t-1}}\right) + \sigma^2$$

The question now is if the iterative system of equations (14) allows convergence at higher values of $\alpha > \alpha_{\max}$. The answer is affirmative, however the maximum load $\alpha$ now depends on the coupling connections, but $\alpha_{\mathrm{coupling}} > \alpha_{\max}$ for any coupling. In particluar, for the coupling of Figure 9 with $a = 0.5$, $\alpha_{\mathrm{coupling}} = 2.81$.

Figure 10 shows that the variance values at the subgraphs at different indexes $t$ evolve completely homogeneously, with a constant iteration delay towards larger $t$. That is, the solutions to the balance equations (14) are linear shifts of each other.

Figure 11 shows the convergence of the variance values at the subgraphs at for an signal-to-noise ratio of 10dB. At that noise level, the system convergence is shown for $\alpha = 1.95$, which is significantly larger than without spatial coupling – see Figure 7.

We may generalize the coupling to include $W$ previous and $W$ future graphs, i.e., utilize a coupling window of $2W+1$, instead of of the basic example in (14). The first $W$ subgraphs are now the anchor symbols and are set to fixed values. The variance update equations at time $t$ can now be computed as

$$x_i^t = \frac{\alpha}{2W+1} \sum_{j=-W}^{W} g\left(\sum_{l=-W}^{W} \frac{1}{2W+1} \frac{1}{x_{i-1}^{t+j+l}}\right) \quad (15)$$

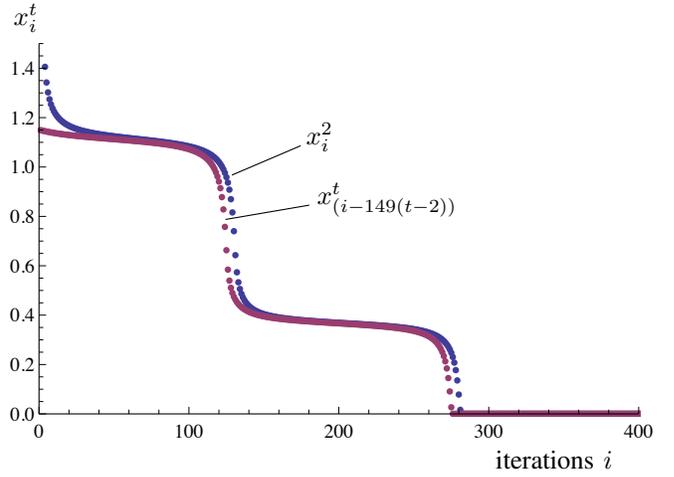

Fig. 10. Homogeneous convergence in the simply coupled graph of Figure 9 with $a = 0.5$ and $\alpha = 2.8$ in the noiseless case.

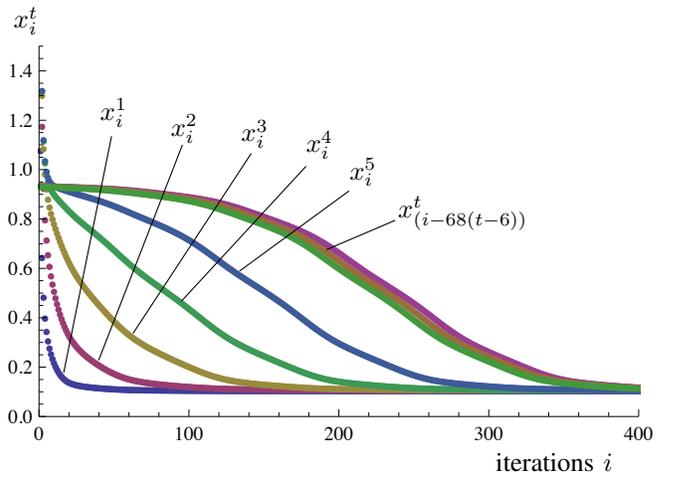

Fig. 11. Convergence in the simply coupled graph of Figure 9 with $a = 0.5$ and $\alpha = 1.95$ in and SNR=10dB, hence the floor at $x_i^t = 0.1$.

This allows us to achieve progressively larger maximum values $\alpha_{\mathrm{coupling}}$ where convergence can be achieved, as substantiated by

*Theorem 1:* For $\sigma^2 = 0$, and with a sufficiently large $W$, i.e., $W = O\left(e^{\alpha \ln \alpha}\right)$, the iteration equations (15) converge to $x_i^t \to 0, \forall t$ and for all values of $\alpha$. In particular we can let $\alpha \to \infty$.

*Proof:* See Appendix I

Theorem 1 shows that the spatially coupled iterative demodulator is no longer "interference" limited, in the sense that it can support an arbitrarily large system load given the necessary signal-to-noise ratio. This is in marked contrast to all known equal-power multi-user detectors, in particular also to the detector in Figure 5. Some values of achievable loads are shown in the table below.

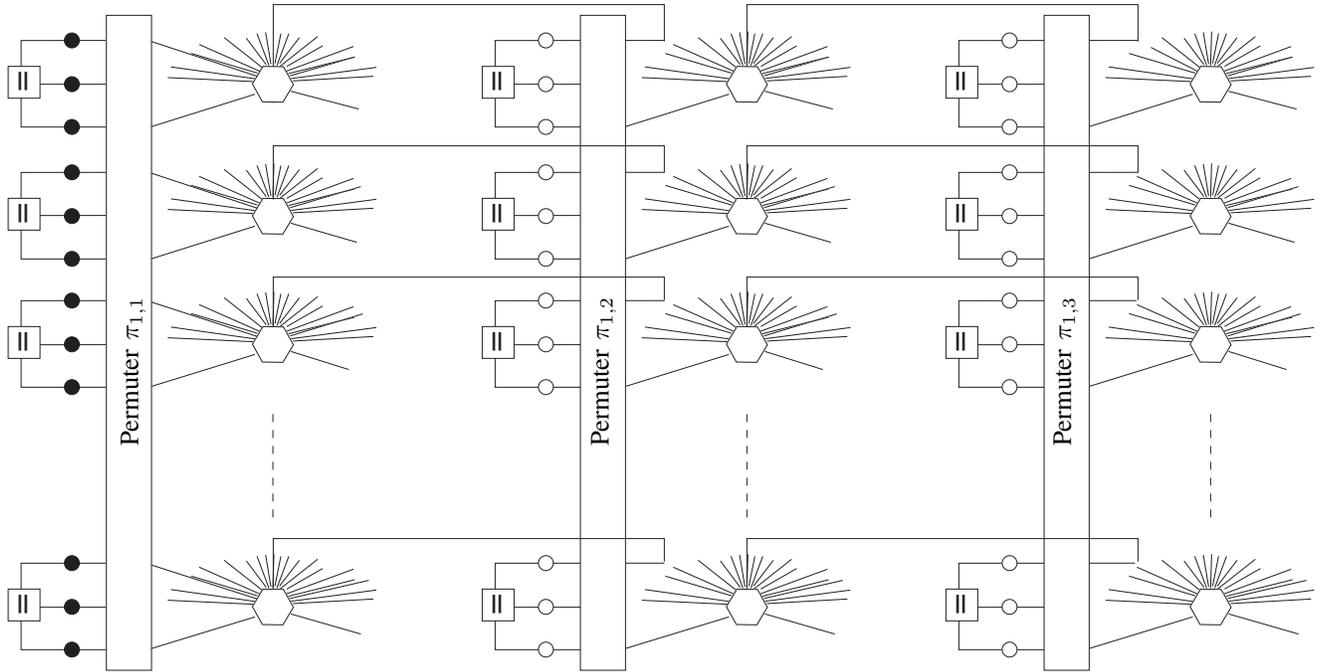

Fig. 9. Basic graphical coupling of different or successive lifted graphs. Couplings with wider span will be more effective.

TABLE I
ACHIEVABLE LOADS AS A FUNCTION OF COUPLING WINDOW SIZE.

| W | 0 | 1 | 2 | 3 | 4 | 5 | 10 | 20 | 50 |
|---|---|---|---|---|---|---|----|----|----|
| $\alpha_{\text{coupling}}$ | 2.07 | 3.17 | 3.6 | 3.9 | 4.1 | 4.3 | 4.9 | 5.5 | 6.2 |

## APPENDIX A
## PROOF OF THEOREM 1

In the spatially coupled system of (15) let $y_i^t$ denote the soft bit variance in the graph replica $t$. Then

$$x_i^t = \frac{\alpha}{2W+1} \sum_{j=-W}^{W} y_i^{t+j} \quad t > 0, i > 0 \tag{16}$$

and the soft bit variance

$$y_i^t = g\left(\frac{1}{2W+1} \sum_{l=-W}^{W} \frac{1}{x_{i-1}^{t+l}}\right) \quad t > 0, i > 0 \tag{17}$$

Combining these two equations we obtain

$$\begin{aligned} x_i^t &= \frac{\alpha}{2W+1} \sum_{j=-W}^{W} y_i^{t+j} \\ &= \frac{\alpha}{2W+1} \sum_{j=-W}^{W} g\left(\frac{1}{2W+1} \sum_{l=-W}^{W} \frac{1}{x_{i-1}^{t+j+l}}\right) + \sigma^2 \end{aligned} \tag{18}$$

We assume that transmission starts at time $t = 1$, i.e., the load of the system increases gradually. At every replica $t$, the rate is increased by $\alpha/(2W+1)$. As a result, the initial condition can be formulated as

$$x_0^t = 0 \quad t \leq 0, \tag{19}$$

$$x_0^t = \frac{\alpha t}{2W+1} + \sigma^2; \quad t \in [1, 2W+1] \tag{20}$$

$$x_0^t = \alpha + \sigma^2 \quad t > 2W+1 \tag{21}$$

and

$$y_0^t = 0 \quad t \leq W \tag{22}$$

$$y_0^t = 1 \quad t \geq W+1 \tag{23}$$

We restate Theorem 1 in a slightly more detailed way for the purpose of this proof.

*Theorem 1:* Consider an interference limited system with $\sigma^2 = 0$, $\alpha > 2$ and

$$W > \exp(15\alpha \ln \alpha + 20\alpha) \tag{24}$$

Then for any $t$

$$\lim_{i \to \infty} x_i^t = 0$$

where after $i \gg 2t$ iterations $x_i^t$ converges to 0 at least exponentially. (The specific numbers 15 and 20 in (24) are merely for convenience).

The proof is broken into two parts. First we show in Lemma 1 that for $W$ satisfying (24) and after $2t$ iterations $x_{2t}^t < \alpha\epsilon$ where

$$\epsilon = \pi \exp\left(-\frac{1}{4\alpha\pi}(2W+1)^{\frac{1}{4\alpha}}\right) \tag{25}$$

In the next step (Lemma 2) we prove that

$$x_{i+1}^t < x_i^t/2$$

for $i > 2t + 2W$. The effect of spatial coupling initiated at position 0 requires $2t$ iterations to propagate to position $t$.

*Lemma 1:* Assume that for $t' < t$ we have noise and interference variance $x_{i-1}^{t'} < \epsilon\alpha$ and for $t' < t+W$ we have soft bit variance $y_{i-1}^{t'} \leq \epsilon$ were $\epsilon$ is given by (25). Then

$$y_{i+2}^{t+W} < \epsilon$$

and as a result $x_{i+2}^t < \alpha\epsilon$.

Note that the initial conditions for spatially coupled system for $t = 1$ are satisfying the conditions of the lemma. Once the lemma is applied to replica $t$ is can be repeated again for replica $t+1$ and so on.

*Proof:* It is easy to see that

$$y_{i-1}^{t'} \leq 1 \quad \text{for } t' \geq t + W$$

According to (16), the above equation, and the condition of the theorem, (namely $y_{i-1}^{t'} \leq \epsilon$ for $t' < t+W$) we obtain

$$x_{i-1}^{t+j} = \frac{\alpha}{2W+1} \sum_{l=-W}^{W} y_{i-1}^{t+j+l}$$
$$\leq \frac{\alpha}{2W+1}((2W-j)\epsilon + j + 1); \; j \in [0, 2W] \quad (26)$$

Let us proceed with iteration $i$. According to (17) and the equation above we obtain

$$y_i^{t+W} = g\left(\frac{1}{2W+1} \sum_{l=-W}^{W} \frac{1}{x_{i-1}^{t+W+l}}\right)$$
$$= g\left(\frac{1}{2W+1} \sum_{j=0}^{2W} \frac{1}{x_{i-1}^{t+j}}\right)$$
$$\leq g\left(\frac{1}{\alpha}\left[\frac{1}{2W\epsilon+1} + \frac{1}{(2W-1)\epsilon+2} + \ldots + \frac{1}{2W+1}\right]\right)$$

and therefore

$$y_i^{t+W} \leq g\left(\frac{1}{\alpha}\frac{1}{2W\epsilon+1}\left[1 + \frac{1}{2} + \ldots + \frac{1}{2W+1}\right]\right)$$

since for any $k > 0$ integer

$$\frac{2W\epsilon+1}{(2W-k)\epsilon+k} \geq \frac{1}{k}$$

For $W$ satisfying (24), and $\epsilon$ as in (25), it can be shown that

$$\epsilon < \frac{1}{2W} \quad (27)$$

and therefore

$$y_i^{t+W} \leq g\left(\frac{\ln(2W+1)}{2\alpha}\right)$$
$$\leq \pi Q\left(\sqrt{\frac{\ln(2W+1)}{2\alpha}}\right)$$
$$\leq \pi \exp\left(-\frac{\ln(2W+1)}{4\alpha}\right)$$
$$\leq \pi(2W+1)^{-\frac{1}{4\alpha}}. \quad (28)$$

Combining (26) and (28) we obtain

$$x_i^t \leq \epsilon\alpha + \frac{\alpha}{2W+1}\pi(2W+1)^{-\frac{1}{4\alpha}}.$$

Likewise for $W$ satisfying (24), and again using (25)

$$\epsilon < \pi(2W+1)^{-1-\frac{1}{4\alpha}} \quad (29)$$

hence

$$x_i^t \leq 2\alpha\pi(2W+1)^{-1-\frac{1}{4\alpha}}.$$

Now

$$y_{i+1}^{t+W} \leq g\left(\frac{1}{2W+1}\frac{1}{2\alpha\pi(2W+1)^{-1-\frac{1}{4\alpha}}} + \ldots\right)$$
$$\leq g\left(\frac{1}{2\alpha\pi(2W+1)^{-\frac{1}{4\alpha}}}\right)$$
$$\leq \pi\exp\left(-\frac{1}{2}\frac{(2W+1)^{\frac{1}{4\alpha}}}{2\alpha\pi}\right) = \epsilon$$

therefore

$$x_{i+1}^t \leq \epsilon(2W+1)\alpha/(2W+1) = \alpha\epsilon.$$

Let $\epsilon_{\text{crit}}$ be the smallest positive root of

$$\exp\left(-\frac{1}{2\alpha\epsilon}\right) = \epsilon/2 \quad (30)$$

Notice that for $\epsilon < \epsilon_{\text{crit}}$

$$\exp\left(-\frac{1}{2\alpha\epsilon}\right) < \epsilon/2, \quad (31)$$

moreover, $\epsilon$ defined by (25) satisfies (31).

*Lemma 2:* Assume that for any $t$ we have $x_{i-1}^t < \epsilon\alpha$ and $y_{i-1}^t \leq \epsilon$. If $\epsilon < \epsilon_{\text{crit}}$ then

$$y_i^t < \epsilon/2$$

and as a result $x_i^t < \alpha\epsilon/2$ for any $t$ as well.

Since $\epsilon$ defined by (25) satisfies (31)

$$x_{i+1}^t < x_i^t/2$$

for $i > 2t+2W$. This means that for $T >> t$, and after $i > T$ iterations $x_i^t$ will decrease to zero exponentially.

*Proof*

$$y_i^t \leq g\left(\frac{1}{2W+1}\left(\frac{1}{\alpha\epsilon} + \ldots + \frac{1}{\alpha\epsilon}\right)\right)$$
$$\leq \exp(-\frac{1}{2\alpha\epsilon}) < \epsilon/2 \quad (32)$$

## APPENDIX B
## RANDOM SIGNAL WAVEFORMS

We are using a set of random unit-length vectors in $n$-space. By random we mean that the vectors are distributed such that their endpoints form a uniform distribution on the $n-1$-dimensional sphere that bounds the $n$-sphere within which these vectors live.

*Prelimaries:*

Consider two vectors $\mathbf{x}$ and $\mathbf{y}$ in 2 dimensions. Then the interference power is given by their crossproduct, i.e.,

$$\rho^2 = (\mathbf{x} \cdot \mathbf{y})^2 = |\mathbf{x}|^2 |\mathbf{y}|^2 \cos^2(\phi)$$

where $\phi$ is the angle between them.

Averaging this over $\phi \in [0, \pi]$ gives 0.5.

*Lemma 3:* The projection power of two random vectors $\mathbf{x}$ and $\mathbf{y}$ in the sense above is given by

$$\overline{\rho^2} = \overline{(\mathbf{x} \cdot \mathbf{y})^2} = \frac{1}{n} \quad (33)$$

*Proof:* Assume without loss of generality that the first basis vector of $n$-space coincides with $\mathbf{x}$. Now decompose $\mathbf{x}$ into two orthogonal parts, $\mathbf{x}_p = \mathbf{x}\cos(\phi)$ and $\mathbf{x}_q = \mathbf{x}\sin(\phi)$. Given that the endpoint of $\mathbf{x}$ is uniformly distributed over the $n-1$ bounding sphere, we first need to compute the ratio of such vectors $\mathbf{x}$ that form the angle $\phi$ with $\mathbf{x}$.

But $\mathbf{y}_q$ is simply a vector of length $\sin(\phi)$ that lives in $n-2$ dimensional sphere of radius $\sin(\phi)$. Therefore, the probability density function of the pair forming an angle $\phi$ is the ratio of the volume of that $n-2$ dimensional sphere to the volume of the $n-1$ dimensional sphere that bounds the space of all vectors. Denoting the volume of an $n$-dimensional unit sphere by $C_n$, the volume of a sphere of radius $R$ is given by $C_n R^n$, and its $n-1$ dimensional surface by $nC_n R^{n-1}$. Therefore, the ratio in question is

$$p(\phi) = \frac{C_{n-1} \sin^{n-2}(\phi)}{nC_n}$$

Given that

$$C_n = \frac{\pi^{n/2}}{\Gamma[n/2+1]}$$

it is a simple exercise to verify the lemma. q.e.d.

Now let us extend the lemma to the case of subspaces of dimension $k > 1$. We assume that $k$ random vectors have been chosen, which span a $k$-dimensional subspace almost surely. We consider the case where a new vector is projected onto that $k$-dimensional subspace (or onto its orthogonal complement). We have the following lemma:

*Lemma 4:* The projection power of a random vector $\mathbf{x}$ onto the orthogonal complement of the space spanned by $k$ random vectors $\mathbf{y}_1, \cdots, \mathbf{y}_k$ is given by

$$\rho_{k,n}^2 = \frac{n-k}{n}$$

*Proof:* Again consider the vector $\mathbf{x}$ decomposed the into two orthogonal parts, $\mathbf{x}_p = \mathbf{x}\cos(\phi)$ and $\mathbf{x}_q = \mathbf{x}\sin(\phi)$, where $\mathbf{x}_p$ lies in the space $\mathcal{S}_k$ spanned by $\mathbf{y}_1, \cdots, \mathbf{y}_k$, and $\mathbf{x}_q$ lies in its complement. The angle $\phi$ is the angle $\mathbf{x}$ forms with $\mathcal{S}_k$. Conditioning on $\phi$, we compute the area on the surface of $C_n$ where the endpoints of $\mathbf{x}$ can lie. The ratio of this area to the area of $C_n$ will give us the probability of $\phi$, since $\mathbf{x}$ is uniformly, and *independently* distributed over the $n$-sphere.

The endpoints of $\mathbf{x}_p$ lie in $\mathcal{S}_k$, in fact the surface of a $k$-dimensional sphere of radius $\cos(\phi)$. The differential volume of that sphere is $kC_k \cos^{k-1}(\phi)$. $\mathbf{x}_q$ is orthogonal to $\mathcal{S}_k$ and lies in the $n-k-1$ dimensional sphere of radius $\sin(\phi)$ (one dimension is fixed because $\mathbf{x}$ has to lie on the surface of the $n$-sphere). The total differential area of possible endpoints for $\mathbf{x}$ is given by

Area of endpoints of $\mathbf{x}$ with $\mathbf{x} \cdot \mathcal{S}_k = \phi$ equals:
$$kC_k \cos^{k-1}(\phi)(n-k)C_{n-k} \sin^{n-k-1}(\phi)$$

The probability that $\mathbf{x}$ forms the angle $\phi$ with $\mathcal{S}_k$ is

$$\Pr(\mathbf{x} \cdot \mathcal{S}_k = \phi) = \frac{kC_k \cos^{k-1}(\phi)(n-k)C_{n-k} \sin^{n-k-1}(\phi)}{nC_n}$$

This is illustrated in the figure below for $n = 3$ and $k = 2$.

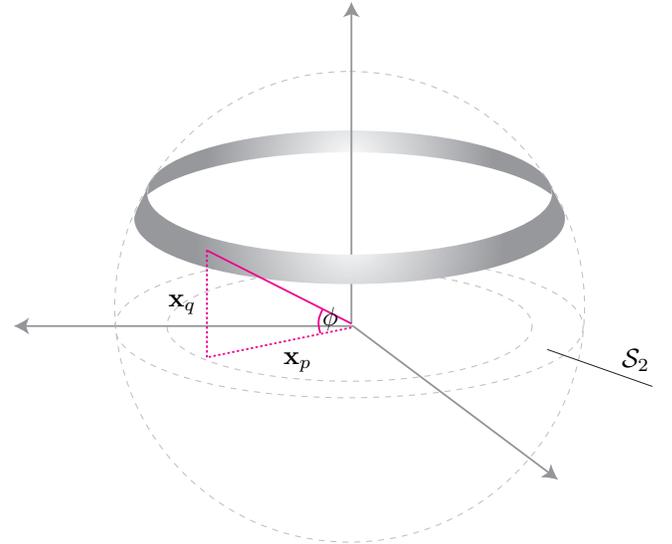

The squared length of the projection $\mathbf{x}_p$, i.e., the power of $\mathbf{x}$ in the subspace $\mathcal{S}_k$ is equal to $\cos^2(\phi)$, and therefore the average projected power of $\mathbf{x}$ onto $\mathcal{S}_k$ is

$$\mathbf{E}\|\mathbf{x}_p\|^2 = \int_0^{\pi/2} \Pr(\mathbf{x} \cdot \mathcal{S}_k = \phi) \cos^2(\phi) d\phi$$

$$= \int_0^{\pi/2} \frac{kC_k \cos^{k+1}(\phi)(n-k)C_{n-k} \sin^{n-k-1}(\phi)}{nC_n} d\phi$$

$$= k/n \quad (34)$$

The last equation can be verified by any of the integration tools. Complementing $\|\mathbf{x}\|^2 = \|\mathbf{x}_p\|^2 + \|\mathbf{x}_q\|^2$, we obtain the lemma. q.e.d.

As observed by the editor, equations (33) and (34) can also be derived using symmetry and isotropy without reference to the geometry used in this appendix.